\documentclass[preprint,aps,floats,showpacs]{revtex4-1}
\usepackage{amssymb}

\usepackage{epsfig}
\usepackage[usenames,dvipsnames]{xcolor}

\newcommand{\be}{\begin{equation}}
\newcommand{\ee}{\end{equation}}
\newcommand{\bea}{\begin{eqnarray}}
\newcommand{\eea}{\end{eqnarray}}

\newcommand{\p}{\partial}
\newcommand{\s}{\sigma}

\newcommand{\la}{\langle}
\newcommand{\ra}{\rangle}
\newcommand{\rd}{\mbox{d}}
\newcommand{\ri}{\mbox{i}}
\newcommand{\re}{\mbox{e}}

\begin{document}
\title{Heisenberg Necklace Model in a Magnetic Field}

\author{A. M. Tsvelik}
\affiliation{Condensed Matter Physics and Material Science Division, Brookhaven National Laboratory, Upton, NY 11973-5000, USA}
\author{I. A. Zaliznyak}
\affiliation{Condensed Matter Physics and Material Science Division, Brookhaven National Laboratory, Upton, NY 11973-5000, USA}

\date{\today}

\begin{abstract}
We study the low-energy sector of the Heisenberg Necklace model. Using the field theory methods, we estimate how the coupling of the electronic spins with the paramagnetic Kondo spins affects the overall spins dynamics and evaluate its dependence on a magnetic field.
We are motivated by the experimental realizations of the spin-1/2 Heisenberg chains in SrCuO$_2$ and Sr$_2$CuO$_3$ cuprates, which remain one-dimensional Luttinger liquids down to temperatures much lower than the in-chain exchange coupling, $J$. We consider the perturbation of the energy spectrum caused by the interaction, $\gamma$, with nuclear spins ($I=3/2$) present on the same sites.
We find that the resulting Necklace model has a characteristic energy scale, $\Lambda \sim J^{1/3}(\gamma I)^{2/3}$, at which the coupling between (nuclear) spins of the necklace and the spins of the Heisenberg chain becomes strong. This energy scale is insensitive to a magnetic field, $B$.
For $\mu_B B > \Lambda$ we find two gapless bosonic modes that have different velocities, whose ratio at strong fields approaches a universal number, $\sqrt 2 +1$. 

\end{abstract}

\maketitle

\section{Introduction}
The problem of interaction of electronic spins with ``extrinsic'' magnetic moments has first emerged in metals, where this interaction leads to strong renormalization of the properties of the bulk \cite{SchriefferWolff1966}.
Quantum critical phases \cite{Sachdev_book} in strongly correlated magnetic insulators whose low-energy behavior is described by a system of spins interacting via Heisenberg exchange, are also sensitive to interactions with extrinsic magnetic moments, such as magnetic impurities, or nuclear spins, which are universally present in real materials \cite{nuclear_spins}.
The response of a spin system near quantum criticality often exhibits fascinating impurity-driven physics, such as found in lightly doped cuprates and related two-dimensional Mott insulators \cite{Sachdev_Science1999,SachdevVojta_PRB2003}. On the other hand, in some complex alloys with rare earths there is a lattice of macroscopically many magnetic rare-earth ``impurity" spins even for ideal stoichiometric materials, and they can dramatically modify the spin dynamics of the host magnetic $3d$ ions \cite{Lynn_PRB1990,Zheludev_PRL1998}. Such cooperative coupling of Ni$^{2+}$ spins to paramagnetic rare-earths induces criticality in the Haldane (S=1) chain antiferromagnet R$_2$BaNiO$_5$, leading to magnetic order at a finite temperature \cite{Zheludev_PRL1998,Zheludev_PRB1998}.

In a spin system near quantum criticality, the hyperfine coupling of the electronic and the nuclear spins becomes important at very low energy, since many abundant magnetic isotopes have non-zero nuclear spin, $I$, \cite{nuclear_spins}. Furthermore, electronic spins can also interact with the nuclei of the nearby ligand ions \cite{AbrahamBleaney}. In studies of magnetic-field-induced quantum phase transitions it has been observed that these hyperfine interactions could result in an avoided electronic quantum criticality. Instead of the expected full softening at the critical field, the electronic spin excitation spectrum was found to have a gap, which increases with the decreasing temperature, while a soft mode is induced in the nuclear spin system, which "takes over" the quantum critical behavior \cite{Ronnow_Science2005,Zaliznyak_JETPL1996,Dumesh_JETP1999}.

Here, we are motivated by the experimental situation  in the chain cuprates, SrCuO$_2$ and Sr$_2$CuO$_3$, probably the most one-dimensional (1D) spin-1/2 model antiferromagnets known to date. These materials have crystal structure composed of chains of corner-sharing CuO$_4$ square plaquettes, where strong Cu-O hybridization results in an exceptionally  strong in-chain superexchange, $J\sim 2500 - 2800$~K, \cite{Zaliznyak_PRL2004,Walters_NatPhys2009}. Small orbital overlaps between the planar CuO$_4$ plaquettes on neighbor chains lead to an extremely small inter-chain coupling, $J'/J \lesssim 5\cdot 10^{-4}$, resulting in an almost ideal spin-chain structure, where a transition to the 3D antiferromagnetically ordered state occurs only below
about 5.5 K \cite{Kojima_PRL1997,Zaliznyak_PRL1999}. Such tremendous disparity between the two energy scales suggests that the hyperfine interactions with Cu nuclear spins ($I=3/2$) might be important. Even though the hyperfine coupling constant for Cu is only $\sim 1.5\cdot 10^{-3}$~meV ($\sim 17$~mK), \cite{AbrahamBleaney} in a combination with the strong in-chain exchange it can generate an energy scale  comparable to the ordering temperature and thus markedly modify  the low-energy dynamics of the 3D ordering.

Recently, a very unusual dependence of the magnon gaps on a magnetic field has been reported in Sr$_2$CuO$_3$, which was attributed to the interaction with the putative Higgs mode \cite{Sergeicheva_2016}. However, a possibility that it could be explained by the necklace-type coupling to the nuclear, or the impurity spins, has not been considered. Our present analysis essentially rules out such a possibility.

The effect of a magnetic field on the system of electronic and nuclear spins coupled by the hyperfine interaction presents an interesting extension of the Kondo chain problem. This is because nuclear magnetic moments are roughly 2000 times smaller than those of electrons, and therefore their interaction with magnetic field can be neglected and field can be considered as acting selectively on the spins of the host antiferromagnetic chain.

We thus consider the case of perhaps the best understood quantum-critical spin system, the one-dimensional spin-1/2 Heisenberg antiferromagnet \cite{book}, where each of the electronic spins is coupled to an additional free spin, which resides on the same lattice site. 
This model, which is an extension of the Kondo necklace model, \cite{Doniach} was previously considered in the context of the Haldane gap problem, where authors tagged it a spin-rotator chain, \cite{Kiselev_PRB2005,Aristov_PRB2010}. The authors established the existence of a characteristic energy scale, $\Lambda$, below which the coupling between the host and the nuclear spins become strong. Our calculations support this and give the same estimate for this scale.

The necklace  model has also been analyzed numerically using the flow equations \cite{Essler_PRB2007}, and authors concluded that quantum critical state only exists at zero coupling, and that any finite coupling to the ``Kondo'' spins generates a spin gap, thus avoiding quantum criticality. We claim that the latter statement is only correct for half-integer nuclear spins; for integer spins, $I$, the model remains critical and in the same universality class as the spin-1/2 Heisenberg chain. We further consider magnetic field, which acts selectively on the host spins, but not on the nuclear ones. It is found that the characteristic energy scale survives and becomes a crossover scale between the weak coupling high energy regime and the strong coupling low energy quantum critical one. At low energy, there are two critical modes with different velocities whose ratio approaches a universal limit at $\mu_B B \gg \Lambda$.

\section{The model}
We consider an isotropic S=1/2 Heisenberg chain where each spin is coupled with another localized spin by an additional isotropic exchange interaction:
\bea
H = \sum_j\Big[ J({\bf S}_j{\bf S}_{j+1}) + \gamma {\bf S}_j{\bf I}_j +\mu_B BS^z_j\Big].
\label{model1}
\eea
As one of the applications this model describes interaction with nuclear spins located on the same sites as the electronic ones, if we adopt an approximation where only contact part of the dipole-dipole interaction is important. In what follows we will call ${\bf I}_j$ ``nuclear''spins although they do not need to be such, as the derivation is carried out for the general case. 

\section{The low energy description. Small magnetic field}
To derive a continuum limit of model (\ref{model1}) it is most convenient to use the path integral representation. In this representation, the nuclear spins are replaced as ${\bf I}_j = I{\bf N}_j$, where ${\bf N}_j$ is a unit vector field with the Berry phase action. As far as the Heiseberg chain is concerned, at energies $\ll J$ we can use the continuum limit description, which is given by the SU$_1$(2) Wess-Zumino-Novikov-Witten (WZNW) theory \cite{affleck} (see also \cite{book}). The resulting action for energies $\ll J$ is given by:
\bea
S = \sum_j I A[{\bf N}_j] + W[g] + \ri\gamma I\sum_j (-1)^j\int \rd \tau {\bf N}_j \mbox{Tr}[\vec\s(g^+-g)],
\label{action}
\eea
where ${\bf I} = I{\bf N}, ~~ {\bf N}^2=1$, $g(\tau,x)$ is the SU(2) matrix field, and $W[g]$ is the action of the SU$_1$(2) WZNW theory, $A[{\bf N}]$ is the Berry phase. The Heisenberg spins are related to the WZNW fields:
\bea
 {\bf S}_j = \frac{\ri}{2\pi}\mbox{Tr}({\vec\s}g\p_x g^+) + \ri(-1)^jC\mbox{Tr}[\vec\s(g-g^+)],
\eea
where $C$ is a nonuniversal amplitude. The WZNW model is a critical theory with a linear excitation spectrum, $\omega = v|k|, ~~v = \pi J/2$.

In the interaction term in (\ref{action}) we kept only the most relevant term, which describes the interaction of the nuclear spins with the staggered magnetization of the Heisenberg chain. This action is not yet what we need since the nuclear spin variables remain discreet. In order to obtain the continuum limit, we have to integrate out the fast components of the nuclear spins. We assume that at low energies the nuclear spins have a short range antiferromagnetic order, so we can write,
\bea
{\bf N}_j = {\bf m}(x) + (-1)^j(1-{\bf m}^2)^{1/2}{\bf n}(x), ~~ x=a_0 j,
\eea
where ${\bf n}^2 =1$ and $|m| \ll 1$. Substituting this into (\ref{action}) and following the well known procedure \cite{haldane} (see also \cite{book}) we obtain,
\bea
&& S = \int \rd\tau\rd x\Big\{ \frac{\ri I}{2}\Big({\bf n}[\p_{\tau}{\bf n}\times\p_x{\bf n}]\Big) + \ri I({\bf m}[{\bf n}\times\p_{\tau}{\bf n}]) \nonumber\\
&& +\ri\gamma I(1-{\bf m}^2)^{1/2}\mbox{Tr}[(\vec\s{\bf n})(g-g^+)]\Big\} + W[g].
\label{S2}
\eea
Now notice that $G = \ri(\vec\s{\bf n})$ is an SU(2) matrix. Hence, $h= gG^+$ is also an SU(2) matrix and we can use the identity \cite{polyakov}:
\bea
W[hG] = W[h] +W[G] +\frac{1}{2\pi}\int \rd\tau\rd x \mbox{Tr}(h^+\p h G\bar\p G^+), \label{identity}
\eea
$\p,\bar\p = \frac{1}{2}(\p_{\tau} \mp \ri v\p_x)$, so that the action (\ref{S2}) becomes
\bea
&& S = S_{mass} + S_{m} + S_{n} + \frac{1}{2\pi}\int \rd\tau\rd x \mbox{Tr}(h^+\p h G\bar\p G^+),\\
&& S_{mass} =  W[h] + \gamma I\int \rd \tau\rd x \mbox{Tr}(h+h^+),\label{h}\\
&& S_m = \int \rd\tau\rd x\Big\{ \frac{D}{2}{\bf m}^2 +\ri I ({\bf m}[{\bf n}\times\p_{\tau}{\bf n}])\Big\},\label{Sm}\\
&& S_n = W[\ri(\vec\s{\bf n})] +I \mbox{(top-term)},\label{Sn}\\
&& \mbox{(top-term)}= \int \rd\tau\rd x\frac{\ri }{2}\Big({\bf n}[\p_{\tau}{\bf n}\times\p_x{\bf n}]\Big) ,
\eea
where
\be
D = \gamma I\la \mbox{Tr}(h+h^+)\ra \sim (I\gamma)^{4/3}.
\label{D}
\ee
The latter estimate follows from the fact that $h$-matrix operator in the SU$_1$(2) WZNW model  has scaling dimension 1/2. In a (1+1)-dimensional critical theory, a relevant perturbation with a scaling dimension $d$ and coupling constant $\lambda$ generates a spectral gap, $\Lambda \sim \lambda^{1/(2-d)}$. Consequently, the perturbation itself acquires a vacuum expectation value, $\sim \Lambda^d \sim \lambda^{d/(2-d)}$, giving rise to (\ref{D}).

Integrating over ${\bf m}$ and taking into account that
\bea
&&W[\ri(\vec\s{\bf n})] = \frac{1}{2\pi}\int \rd\tau\rd x[v^{-1}(\p_{\tau}{\bf n})^2 +v(\p_x{\bf n})^2] \nonumber\\
&& + (1/2)\times \mbox{(top-term)},
\eea
we obtain the effective Lagrangian density for the slow field ${\bf n}$:
\bea
&&{\cal L} = \frac{1}{2}\Big(\frac{I^2}{D} + \frac{1}{\pi v}\Big)(\p_{\tau}{\bf n})^2 + \frac{v}{2\pi}(\p_x{\bf n})^2 \nonumber\\
&& + \frac{\ri (I-1/2)}{2}\Big({\bf n}[\p_{\tau}{\bf n}\times\p_x{\bf n}]\Big) ,
\label{sigma}
\eea
plus the action for the massive part:
\bea
&&S_{mass} = W[h] +\gamma I\int \rd \tau\rd x \mbox{Tr}(h+h^+) \nonumber\\
&& + \int \rd\tau\rd x \mbox{Tr}({\bf J}_L[{\bf n}\times\bar\p {\bf n}]).
\label{mass1}
\eea
The mass gap $\Lambda$ serves as the ultraviolet cut-off for the sigma model (\ref{sigma}).  The corrections to the sigma model generated by the last term in (\ref{mass1}) carry higher power of gradients of the ${\bf n}$-field and therefore can be discarded for momenta $< \Lambda v^{-1} $.

The uniform magnetic field couples only to the electronic spins, through WZNW currents:
\bea
&&\mu_B B(J_L^z + J_R^z) = \frac{\mu_BB}{2\pi} \mbox{Tr}\Big[\s^z (g\bar\p g^+ - g\p g^+)\Big] \nonumber\\
&& = \frac{\ri\mu_BB}{2\pi}\mbox{Tr}\Big(\s^zg\p_x g^+\Big).
\eea
After the transformation, $g = hG$, we obtain,
\bea
\ri[h\s^zh^+ (G\p_x G^+) + \s^z h\p_x h^+] \approx \ri\s^z[(G\p_x G^+) +  h\p_x h^+],
\eea
that is, the field couples uniformly to the high and low energy modes.

\section{Extracting the results}
Let us take a closer look at (\ref{mass1}). We will use the remarkable property of the SU$_1$(2) WZNW that it can be reformulated as a theory of noninteracting bosonic field (the Gaussian model) so that
\bea
&&W[h] +\gamma I\int \rd \tau\rd x \mbox{Tr}(h+h^+) \nonumber\\
&& = \int \rd\tau \rd x\Big[ \frac{1}{2}(\p_{\mu}\Phi)^2 - \gamma I\cos(\sqrt{2\pi}\Phi)\Big],
\eea
with the identification,
\bea
h= \frac{1}{\sqrt{2}}\left(
\begin{array}{cc}
\re^{\ri\sqrt{2\pi}\Phi} & \re^{\ri\sqrt{2\pi}\Theta}\\
-\re^{-\ri\sqrt{2\pi}\Theta} & \re^{-\ri\sqrt{2\pi}\Phi}
\end{array}
\right),
\label{matrix}
\eea
where $\Theta$ is the field dual to $\Phi$: $[\Phi(x), \p_y\Theta(y)] = \ri\delta(x-y)$.
Hence, the model (\ref{h}) describing the high energy part of the spectrum is equivalent to the sine-Gordon model with a coupling constant $\beta^2 =2\pi$:
\bea
&&{\cal L}_{mass} = \frac{1}{2}(\p_{\mu}\Phi)^2 - \gamma I(1-<{\bf m}^2>)^{1/2}\cos(\sqrt{2\pi}\Phi) \nonumber\\
&& + \frac{\mu_BB}{\sqrt\pi}\p_x\Phi,
\eea
 and is exactly solvable (see, for example, \cite{zam} or \cite{book}). The spectrum at this value of $\beta$ in zero magnetic field consists of a massive triplets and one singlet breather  with the mass $\Lambda_2 = \sqrt 3 \Lambda_1$:
\be
\epsilon(p)_n = \sqrt{(vp)^2 + \Lambda_n^2}, ~~ \Lambda_n \sim J^{1/3}(\gamma I)^{2/3}(1-m_z^2)^{1/3}.
\label{mass}
\ee
Here $\Lambda_1 = C\Lambda$ with $C$ being a known numerical constant.

A finite magnetic field has two effects. Firstly, it creates finite magnetization of the nuclear spins, $\langle m^z\rangle \neq 0$, and this leads to a decrease in the mass gap (\ref{mass}). Secondly, as we discuss below, for fields above the threshold, $\mu_B B > \Lambda_1$, the magnetization appears in the Heisenberg spin sector. 

The sigma model (\ref{sigma}) can be conveniently rewritten in the canonical form,
\bea
&&{\cal L} = \frac{1}{2\pi g}[c^{-1}(\p_{\tau}{\bf n})^2 +c(\p_x{\bf n}-({\bf h}\times{\bf n}])^2] \nonumber\\
&& + \ri (I-1/2)\times\mbox{(top-term)}, ~~ {\bf n}^2 =1,
\label{Sigma2}
\eea
where $g$ is the dimensionless coupling constant,
\be
c = vg, ~~g = (1+\pi vI^2/D)^{-1/2}\sim (\gamma/J\sqrt I)^{2/3}.
\ee
The condition $g \ll 1$ validates the self-consistency of the semiclassical approach adopted here. The O(3) sigma model with the topological term is integrable when the coefficient $I$ at this term is an integer factor \cite{zam,wieg,zamfat}. If $I-1/2$ is half integer, the spectrum of the sigma model (\ref{Sigma2}) is gapless, if not, it is gapped:
\bea
\epsilon(p) =\sqrt{c^2p^2 + \Delta^2}, ~~ \Delta \sim \Lambda g^{-1}\re^{-1/g},
\eea
where $\Lambda \sim J^{1/3}(\gamma I)^{2/3}$ is the sigma model energy cut-off.

The above derivation is valid for any type of necklace, but the case most interesting from the practical point of view is when the $I$-spins are nuclear ones not just in name. For nuclear spins the exchange coupling  $\gamma$ is minuscule, and therefore the sigma model coupling constant is also tiny. As a result, one can neglect all nonlinearities in (\ref{Sigma2}) and consider ${\bf n}$ as a free field. Hence, for magnetic fields small in comparison to $\Lambda$ the low-energy spectrum of the necklace model consists of three gapless modes:
\bea
\omega^z = c|q|, ~~ \omega^{x,y} =c |q- h|.
\eea

\section{Strong magnetic field $\mu_BB \gg J^{1/3}(I\gamma)^{2/3}$}
Here we consider the situation where the magnetization caused by the applied field is small, $\la S^z\ra \ll1$. In this case one can neglect the changes in the Luttinger parameter caused by the field. The uniform field can be removed by a unitary transformation, so that the longitudinal part of the staggered magnetization becomes oscillatory. Then, it is more convenient to use the abelian bosonization:
\bea
S^z - m = \frac{1}{\sqrt{\pi}}\p_x\Phi + (-1)^n\sin(\sqrt{2\pi}\Phi + 2k_Fx)+...
\eea
The staggered components of the transverse magnetization are $\cos(\sqrt{2\pi}\Theta)$ and $\sin(\sqrt{2\pi}\Theta)$, where the dual field $\Theta$ is not shifted.

In order to maximize the energy gain, the nuclear spins have to adjust to the Heisenberg spins-1/2. The best configuration still respecting the condition ${\bf I}_j^2= I^2$ is,
\bea
&& {\bf I}_j/I ={\bf m} + \sqrt{1-{\bf m}^2} (-1)^j{\bf n},\\
&& {\bf n} = (\cos(\alpha + 2k_Fx), \sin(\alpha +2k_Fx)\cos(2k_Fx+\phi), \nonumber\\
&& \sin(\alpha +2k_Fx)\sin(\phi +2k_Fx)),
\label{caln}
\eea
where $\p_x\alpha, \p_x\psi \ll 2k_F$ and ${\bf m}$ is a fluctuational component to be integrated out.
Discarding the fast oscillatory terms, we obtain for the exchange interaction:
\bea
&& \frac{\gamma \sqrt{1-{\bf m}^2}}{2}[\sin(\sqrt{2\pi}\Phi - \alpha) + \sin(\alpha -\psi - \sqrt{2\pi}\Theta)] \nonumber\\
&& = \frac{\gamma \sqrt{1-{\bf m}^2}}{ 2}\mbox{Tr}(h^+g + g^+h),
\eea
where
\bea
g^+ = \frac{\ri}{\sqrt 2}\left(
\begin{array}{cc}
-\re^{-\ri\alpha} & \re^{\ri(\alpha-\psi)}\\
\re^{\ri(\psi-\alpha)} & \re^{\ri\alpha}
\end{array}
\right), ~~h= \frac{1}{\sqrt{2}}\left(
\begin{array}{cc}
\re^{\ri\sqrt{2\pi}\Phi} & \re^{\ri\sqrt{2\pi}\Theta}\\
-\re^{-\ri\sqrt{2\pi}\Theta} & \re^{-\ri\sqrt{2\pi}\Phi}
\end{array}
\right).
\eea
Now we can use the fact that the XXX spin-1/2 Heisenberg model is described by the SU$_1$(2) WZNW model with matrix field $h$. We shift $h$-field by $g$ using the identity (\ref{identity}).
This yields the massive sine-Gordon theory and the WZNW action:
\bea
W[h] =\frac{1}{16\pi} \int \rd\tau \rd x [(\p_{\mu}\alpha)^2 + (\p_{\mu}(\psi-\alpha))^2] \label{w}
\eea
The Berry phase is still equal to
\be
\ri I\int \rd\tau \rd x \Big({\bf m}[{\bf n}\times\p_{\tau}{\bf n}]\Big),
\ee
where ${\bf n}$ in this case is given by (\ref{caln}).
We have to integrate it over ${\bf m}$, as in Eq. \ref{Sm}, to get,
\bea
&& \int \rd x  \frac{1}{2D}(\p_{\tau}{\bf n})^2 = \frac{1}{2D}\int \rd x \Big[(\p_{\tau}\alpha)^2 + \sin^2(2k_Fx +\alpha)(\p_{\tau}\psi)^2\Big] \nonumber\\
&& = \frac{1}{2D}\int \rd x \Big[(\p_{\tau}\alpha)^2 + \frac{1}{2}(\p_{\tau}\psi)^2\Big] ,
\label{berry}
\eea
where $D$ is given by (\ref{D}). Combining (\ref{w},\ref{berry}) we get the effective low-energy description in terms of the Gaussian model of two fields
\bea
&& {\cal L} = \frac{1}{2}(D^{-1} + 1/4\pi v)\Big[(\p_{\tau}\alpha)^2 + \frac{1}{2}(\p_{\tau}\psi)^2\Big] + \frac{v}{8\pi}\Big[(\p_x\alpha)^2 \nonumber\\
&& + \frac{1}{2}(\p_x\psi)^2\Big] - \frac{1}{8\pi}\Big[v^{-1}\p_{\tau}\alpha\p_{\tau}\psi + v\p_x\alpha\p_x\psi\Big].
\eea
Diagonalizing the action in the leading order in $D/v$ we obtain the following spectrum:
\bea
\omega_{\pm} = |q|\Big[\frac{Dv}{4\pi}(1\pm 1/\sqrt 2)\Big]^{1/2}
\eea
There are two gapless modes whose ratio of the velocities is equal to $\sqrt{2} + 1$. They exist at momenta $|q| < \Lambda v^{-1}$, below the gap of the sine-Gordon model. The energy scale of the latter is impervious to the magnetic field and, as in zero field, it marks a crossover from the dynamics dominated by the Heisenberg spins to the strongly coupled dynamics.




\section{Conclusions}
The summary of the results for zero magnetic field is as follows. We have demonstrated that the Necklace model (\ref{model1}) has a characteristic energy scale, $\Lambda \sim J^{1/3}(\gamma I)^{2/3}$, at which the coupling between spins of the necklace and the spins of the Heisenberg chain becomes strong. This energy scale may be quite sizable even when the necklace is made of nuclear spins, provided the intrachain exchange is large. For the experimental case of Sr$_2$CuO$_3$, where $I=3/2$, $J \approx 250$~meV ($\approx 2800$~K) and $\gamma \sim 1.5\cdot 10^{-3}$~meV ($\sim 17$~mK), \cite{AbrahamBleaney} we obtain $\Lambda \sim 0.1$~meV ($\sim 1$~K). This new energy scale is of the same order of magnitude as the Neel temperature, $T_N \approx 5.5$~K, and therefore interactions with nuclear spins can be expected to contribute significantly to the magnon dynamics at low temperature, $T \ll T_N$.

Thus, at high energies compared to the cutoff, $\Lambda$, we obtain the weak coupling limit of the Heisenberg chain and the Kondo spins; the chain spectrum is linear, with velocity $v = \pi J/2$. At energies below $\Lambda$, the dynamics is described by the O(3) nonlinear sigma model with the topological term (\ref{sigma}). Strong retardation effects coming from the dynamical interaction between the nuclear spins and the critical fluctuations of the Heisenberg chain strongly renormalize the coupling constant $g$ and the asymptotic velocity, $c= gv$, of the sigma model excitations. Its spectrum remains gapless if the nuclear spins are integer and acquires an exponentially small gap otherwise. 

The energy scale $\Lambda$ survives in a strong magnetic field, but the character of the low energy sector changes. For  $\mu_B B > \Lambda$, the dynamics is described by two gapless bosonic modes with different velocities. Their ratio at strong fields approaches a universal number, $\sqrt 2 +1$. The high-energy spectrum above the cutoff, $\Lambda$, is insensitive to magnetic field that is strong compared to $\Lambda$, but small compared to the in-chain Heisenberg exchange coupling, $J \gg \mu_B B > J^{1/3}(\gamma I)^{2/3}$ (i.e. when the magnetization is small). Therefore, this interaction cannot explain an unusually strong magnetic field dependence of the magnon gaps in Sr$_2$CuO$_3$ observed by ESR \cite{Sergeicheva_2016}, which authors have attributed to magnon interaction with the putative Higgs mode.

\begin{acknowledgments}
\noindent
This work was supported by the Office of Basic Energy Sciences (BES), Division of Materials Sciences and Engineering, U.S. Department of Energy (DOE), under Contract No. DE-SC00112704.
\end{acknowledgments}

\end{document}